\documentclass[twocolumn,preprintnumbers,prl,amsmath,amssymb,superscriptaddress,showpacs]{revtex4}
\usepackage{epsfig}
\usepackage{graphicx}

\begin{document}
\title{Experimental quantum key distribution over highly noisy channels}

\author{L.-P. Lamoureux}
\affiliation{Quantum Information and Communication, Ecole
Polytechnique, CP 165/59, Universit\'e Libre de Bruxelles, 
Avenue F. D. Roosevelt 50, 1050 Bruxelles, Belgium}

\author{E. Brainis}
\affiliation{Optique et Acoustique, CP 194/5, Universit\'e Libre
de Bruxelles, Avenue F. D. Roosevelt 50, 1050 Bruxelles, Belgium}

\author{N. J. Cerf}
\affiliation{Quantum Information and Communication, Ecole
Polytechnique, CP 165/59, Universit\'e Libre de Bruxelles, 
Avenue F. D. Roosevelt 50, 1050 Bruxelles, Belgium}

\author{Ph. Emplit}
\affiliation{Optique et Acoustique, CP 194/5, Universit\'e Libre
de Bruxelles, Avenue F. D. Roosevelt 50, 1050 Bruxelles, Belgium}

\author{M. Haelterman}
\affiliation{Optique et Acoustique, CP 194/5, Universit\'e Libre
de Bruxelles, Avenue F. D. Roosevelt 50, 1050 Bruxelles, Belgium}

\author{S. Massar}
\affiliation{Service de Physique Th\'eorique, CP 225, Universit\'e
Libre de Bruxelles, Boulevard du Triomphe, 1050 Bruxelles, Belgium}
\affiliation{Quantum Information and Communication, Ecole
Polytechnique, CP 165/59, Universit\'e Libre de Bruxelles, 
Avenue F. D. Roosevelt 50, 1050 Bruxelles, Belgium}

\begin{abstract}
Error filtration is a method for encoding the quantum state 
of a single particle into a higher dimensional Hilbert
space in such a way that it becomes less sensitive to phase noise.
We experimentally demonstrate this method by distributing a secret key over
an optical fiber whose noise level otherwise precludes secure quantum 
key distribution.
By filtering out the phase noise, a bit error rate of 15.3\% $\pm$ 0.1\%, 
which is beyond the security limit, can be reduced to 10.6\% $\pm$ 0.1\%,
thereby guaranteeing the cryptographic security.
\end{abstract}

\maketitle

One of the central results in quantum information is that errors
can, in principle, be corrected \cite{S,BBPSSW}. This discovery
transformed quantum information from an intellectual game into a
field which could revolutionize the way we process information.
Practical realizations of quantum error correcting codes 
are, however, extremely difficult because they require multiparticle interactions. 
A first experimental demonstration of quantum error correction has recently 
been realized \cite{PGUWZ}, but it is still very far from being usable
in practical applications. An alternative method, called \emph{error filtration}, allows errors to be filtered out during quantum
communication, and can in contrast be easily implemented using present technology \cite{GLMP}. The main idea of
error filtration is to encode one qubit in a single particle within a Hilbert space of dimension greater than
two. It is then possible to detect, with high probability, whether 
a phase error has occurred, and, if so, to discard the state. This quantum error detection scheme is less powerful than full error correction, but, for many
applications such as quantum key distribution (QKD), discarding the state affected by phase noise is sufficient. The advantage of this method is that the encoding and decoding operations do not require multiparticle interactions, hence they are relatively easy to implement
by interferometric techniques.
\par

Here, we report on an experimental demonstration of error filtration. For a detailed theoretical
introduction to the method, we refer to \cite{GLMP} where it is also shown that error filtration can be extended to the purification of
entanglement affected by phase noise.
Our experimental scheme is based on the fiber optics ``plug
and play" quantum cryptosystem \cite{Retal,SGGRZ} and its
extension to more than two dimensions presented in \cite{BLCEHM} in
the context of quantum computing. The interest of error
filtration is illustrated by performing QKD over a noisy quantum 
communication channel with so much noise that a secure BB84 protocol 
cannot be realized. It is known that if the bit error rate (BER) exceeds 14.6\%,
then a simple cloning attack makes the BB84 protocol insecure. On the other hand, if the BER is lower than 11.0\%, then the BB84 protocol is provably
secure. Between the two boundaries lies a gray zone where the
security of BB84 is unknown, see \cite{GRTZ} for a review. In our
experiment, we consider an error prone QKD scheme 
with a BER = 15.3\% $\pm$ 0.1\%, which is therefore insecure. 
Using error filtration, the BER
is brought down to 10.6\% $\pm$ 0.1\% so that QKD is rendered
secure. Nevertheless, the present work will probably not have
immediate practical applications in QKD because the main
limitation at present is absorption rather than phase noise. In order to
extend QKD much beyond the present limit of 100~km, new techniques
such as quantum repeaters will have to be used (see
\cite{DLCZ,GMRTZ} for recent proposals). Then, the
contribution of phase noise will probably become significant
and error filtration could provide a simple solution. Finally, note
that it was realized previously that the use of higher dimensional
systems can increase the resistance of QKD to noise \cite{CBKG}.
We will discuss the advantages of error filtration over these previous
methods in what follows.
\par

\begin{figure*}
\centerline{\includegraphics[width=0.9\linewidth]{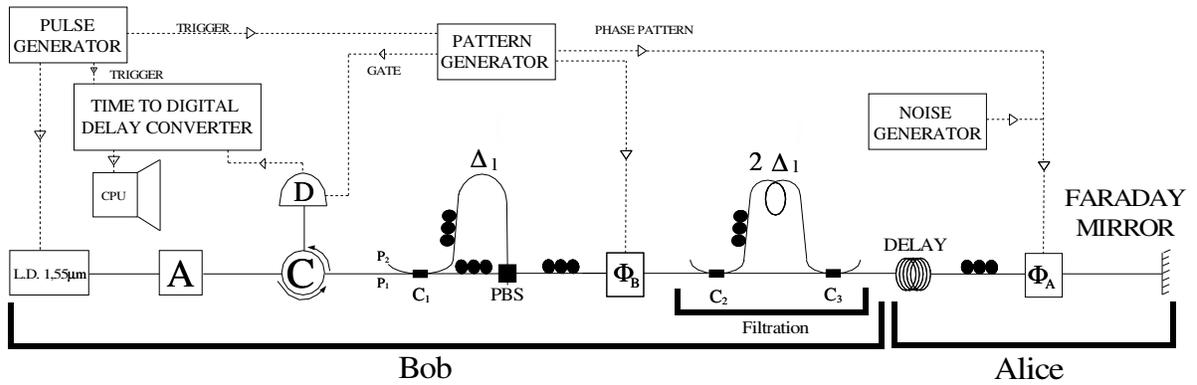}}
\caption{Fiber optics QKD setup with error filtration.} \label{fig:gfx}
\end{figure*}

Our experimental error filtration setup 
works with attenuated coherent states traveling in localized time
bins in optical fibers (standard SMF-28), see Fig.~\ref{fig:gfx}.
Bob uses a laser diode at 1.55 $\mu$m to produce a 3-ns light
pulse. The pulse is attenuated by an optical attenuator (Agilent
8156A), and then split by a first 50/50 coupler (C$_1$) to produce
two pulses. These pulses impinge on the input ports of a polarization
beamsplitter (PBS) with a time delay $\Delta = 60$ ns.
Their polarizations are rotated, using polarization
controllers, so that the two pulses exit by the same port of the
PBS. Next, the two pulses are processed through an unbalanced
Mach-Zehnder (MZ) interferometer C$_2$C$_3$ with a path length
difference equivalent to 2$\Delta$. This produces four emerging
pulses traveling down a 900-ns delay line (representing the noisy
communication channel). Then, at Alice's site, the pulses are
reflected back by a Faraday mirror. The use of a Faraday mirror
makes the setup insensitive to birefringence in the delay line.
Alice may modify the phases of the four pulses using a phase
modulator ($\phi_A$) (Trilink) controlled by a pattern generator
(Agilent 81110A). When they reach Bob's site, the pulses pass through 
the MZ interferometer and interfere at the C$_2$ coupler, which, 
as we shall show, realizes error filtration. The six emerging
pulses then travel trough Bob's phase modulator ($\phi_B$) (Trilink)
and the PBS to the coupler C$_1$, where they interfere and are sent
to a single-photon detector (id Quantique id200) via a circulator.
The detector was gated by a pattern generator during a 5-ns window around
the arrival time of the central pulse. Its output was registered
using a time-to-digital delay converter (ACAM-GP1) connected to a
computer. All electronic components were triggered by a pulse
generator (Standford Research Inc. DG355). In order to maximize
the interference visibilities, polarization controllers were
introduced in the long arm of the MZ interferometer and in front
of the polarization-sensitive phase modulators. Once optimized,
the setup was stable for days.
\par

Before discussing error filtration, let us 
first consider the case where the MZ interferometer
C$_2$C$_3$ is absent. Then our setup is identical to the
plug-and-play quantum cryptography setup \cite{Retal}. The
wavefunction describing the state after Alice has encoded her
phase $\phi_A$ can be written as $ {1\over\sqrt{2}}(|t_0\rangle_H
- e^{i\phi_A} |t_1\rangle_V)$, where the subscripts $H$ and $V$
represent the polarization states while the
subscripts 0 and 1 represent the relative delay of time bin $i$,
i.e. $t_i = i\cdot\Delta$. In the BB84 protocol, the phase $\phi_A$
must be chosen randomly in $\{0,{\pi\over{2}},\pi,{3\pi\over{2}}\}$. 
The ($-1$) phase in front of 
$|t_1\rangle_V$ takes into account the conventional relative phase
of $\pi/2$ between  the reflected and transmitted light at coupler
C$_1$ and at the PBS. The state leaving Alice's site thus reads
\begin{equation}
|\psi\rangle={1\over\sqrt{2}}\left[ |t_0\rangle_V - e^{i\phi_A} |t_1\rangle_H \right].
\end{equation}
where the polarizations have been interchanged because of the Faraday mirror.
\par

Intrinsic noise in the plug-and-play QKD scheme is actually very
low, see \cite{SGGRZ}, so that in our experiment we had to simulate 
the noisy channel by making Alice's phase modulator 
imperfect. This was achieved by electronically combining the signal 
from the pattern generator with the output of a function generator (Agilent
33250A) producing gaussian electronic noise with an adjustable
amplitude and a 50-MHz 3dB-bandwidth. Since the time bins are
separated by 60~ns, the phase noise in the successive time bins can be
considered as independent. The state produced by Alice is thus
\begin{equation}
{1\over{\sqrt{2}}}\left[e^{i\varphi_0}|t_0\rangle_V-e^{i(\phi_A+\varphi_1)} |t_1\rangle_H\right].
\label{state2d}
\end{equation}
where the $\varphi_i$'s are independent random phases 
drawn from a gaussian probability distribution
$P(\varphi_i)={1\over{\sqrt{2\pi\sigma^2}}} \exp\left(-\varphi_i^2/(2\sigma^2)\right)$ of tunable
variance $\sigma^2$. The density matrix obtained by averaging over the random phases is $\rho = e^{-\sigma^2}
|\psi\rangle\langle \psi| + (1 - e^{-\sigma^2})\openone/2$, so
that phase noise amounts to the admixture of isotropic noise.
In terms of bit error rate, this noise level thus corresponds to
$\mathrm{BER}=1-\langle\psi|\rho|\psi\rangle = (1 - e^{-\sigma^2})/2$.
\par

\begin{figure}[htb]
\centerline{\includegraphics[width=0.9\linewidth]{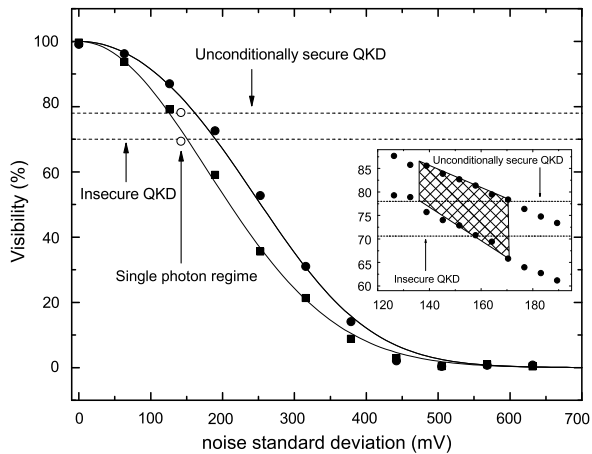}}
\caption{Visibility as a function of the standard deviation 
of the noise signal produced by the function generator.  
The proportionality factor between the x-axis and $\sigma$ 
was determined by fitting the experimental data 
without filtration with Eq. (\ref{V2}). 
The squares (full circles) represent the measured
visibilities without (with) filtration, while the curves are the
theoretical predictions of Eqs. (\ref{V2})
and (\ref{V4}). The open circles show filtration in the
single-photon regime. The inset magnifies
the data in the region where one passes from
an insecure ($V\le 70.7$\%) to a secure ($V>78.0$\%) QKD protocol.
The criss-crossed area shows that
a visibility that is either insecure or of unknown security
can be increased to a provably secure one. By making the statistics
over 200,000 runs, the error bars were made
smaller than the plotting points. } \label{fig2}
\end{figure}

The two pulses travel back to Bob, who performs a
measurement in the ${1\over{\sqrt{2}}}(|t_0\rangle_V\mp
e^{-i\phi_B}|t_1\rangle_H)$ basis by applying the phase $\phi_B\in
\{0,\frac{\pi}{2}\}$. Bob's detector is connected to either output
port P$_1$ or P$_2$. The probability for Bob to detect a count
(assuming a perfect detector) is 
${1\over2}\pm{1\over2}\cos(\varphi_1-\varphi_0 +\phi_A + \phi_B)$
where the sign depends on which output port is used. The
visibility of the interference fringes measured by Bob is
\begin{equation}
V = { I_{\rm max} - I_{\rm min} \over I_{\rm max} + I_{\rm min}} 
= {e^{-\sigma^2}} . 
\label{V2}
\end{equation}
The measured values of $V$ agree well with this prediction, see 
lower curve in Fig.~\ref{fig2}. Note that the visibility is related to
the bit error rate via $\mathrm{BER}= (1-V)/2$.
\par

In order to carry out error filtration, we use the additional MZ
interferometer C$_2$C$_3$. Each pulse exiting from the PBS is
split into two pulses by the MZ interferometer according to
$|t_0\rangle_H \to (|t_0\rangle_H - |t_2\rangle_H)/2$,
$|t_1\rangle_V \to (|t_1\rangle_V - |t_3\rangle_V)/2$.
The factor $1/2$ takes into account that half of the intensity is
lost at coupler C$_3$. After Alice has encoded her phase $\phi_A$
and the pulses have been reflected by the Faraday mirror, the
state becomes
\begin{equation}
{{1}\over{2\sqrt{2}}}\left[{|t_0\rangle_V - |t_2\rangle_V -
e^{i\phi_A}(|t_1\rangle_H - |t_3\rangle_H)}\right]. \label{seven}
\end{equation}
This is formally identical to the BB84 protocol since Alice
effectively uses the 2-dimensional space ${\cal S}$ 
spanned by $(|t_0\rangle_V - |t_2\rangle_V)$ and $(|t_1\rangle_H
-|t_3\rangle_H)$. The time bins $t_2$ and $t_3$ are just
replica of the time bins $t_0$ and $t_1$, respectively. 
Then, because of the noise, the four pulses get random phases 
$\varphi_0$, $\varphi_1$,
$\varphi_2$ and $\varphi_3$, respectively. The state that Alice
sends back to Bob is thus
\begin{equation}
{1\over{2\sqrt{2}}} \left[ e^{i\varphi_0}|t_0\rangle_V -
e^{i\varphi_2}|t_2\rangle_V -
e^{i\phi_A}(e^{i\varphi_1}|t_1\rangle_H-e^{i\varphi_3}|t_3\rangle_H)\right] .
\end{equation}
It now belongs to the full space spanned by $|t_0\rangle_V$,
$|t_1\rangle_V$, $|t_3\rangle_H$ and $|t_4\rangle_H$, since the phase noise
has taken it out of the space ${\cal S}$.
The idea of error filtration is for Bob
to project the state back onto ${\cal S}$ in order to selectively enhance
the visibility. This projection is realized by the MZ interferometer.
It succeeds with probability $1/2$ in our implementation because
at coupler C$_3$, each pulse only has a probability $1/2$ of
taking the right path \footnote{This could, in
principle, be remedied by replacing the coupler C$_3$ by an optical
switch which directs the light pulses along the appropriate
paths. High speed, low-loss optical switches are not available
commercially at present, but this is a technological 
rather than fundamental limitation.}. When time bins
$t_0$ and $t_1$ follow the long path in the MZ interferometer while
time bins $t_2$ and $t_3$ follow the short path, the time bins
$t_0$ and $t_1$ then interfere respectively with $t_2$ and $t_3$ at
coupler C$_2$ and error filtration occurs. The component of the
resulting quantum state into the time window $(t_2,t_3)$
is the 2-dimensional state
\begin{equation}\label{eqn6}
\frac{-1}{4\sqrt{2}}\left[(e^{i\varphi_0}+e^{i\varphi_2})|t_2\rangle_V
-e^{i\phi_A}(e^{i\varphi_1}+e^{i\varphi_3})|t_3\rangle_H\right],
\end{equation}
where the normalization takes into account the losses at the
unconnected pigtail at coupler C$_2$. The interference of phase factors 
in the amplitudes is responsible for the error filtration.
All other optical paths
do not participate to error filtration, as we can see by considering the full state entering Bob's phase modulator
\begin{eqnarray}
{1\over{4\sqrt{2}}}\left[ e^{i\varphi_0}|t_0\rangle_V -
e^{i(\phi_A + \varphi_1)}|t_1\rangle_H -(e^{i\varphi_0}+
e^{i\varphi_2})|t_2\rangle_V \right.
\nonumber\\
\left. +e^{i \phi_A}(e^{i\varphi_1}+e^{i\varphi_3})|t_3\rangle_H
+e^{i\varphi_2}|t_4\rangle_V-
e^{i(\phi_A+\varphi_3)}|t_5\rangle_H)\right].
\end{eqnarray}
Indeed, the time bins $(t_0,t_1,t_4,t_5)$ contain information about
$\phi_A$, but phase errors are not filtered out
since $(t_0,t_1)$ and $(t_4,t_5)$ have the same structure
as Eq.~(\ref{state2d}).
\par

Bob completes his measurement by encoding the phase $\phi_B$ and
letting the time bins interfere at coupler C$_1$. In the time window $(t_2,t_3)$, this realizes a
measurement in the ${1\over{\sqrt{2}}}(|t_2\rangle_V\mp
e^{-i\phi_B}|t_3\rangle_H)$ basis. In fact,
there will be three time bins $t_0'$, $t_2'$, and $t_4'$ 
at the output of coupler C$_1$. Time bin $t_2'$ corresponds
to the measurement of the filtered state, Eq. (\ref{eqn6}), whereas 
time bins $t_0'$ and $t_4'$ correspond to the measurement of the
unfiltered time bins $(t_0,t_1)$ and $(t_4,t_5)$, respectively.
Assuming that the noise affecting the different time bins is
independent, the visibility of the interference fringes in the
time bin $t_2'$ is
\begin{equation}
V={2 / ( 1 + e^{\sigma^2})}. \label{V4}
\end{equation}
Thus, filtration effects an increase of visibility at the
expense of a decreased intensity (the information is lost when the photon ends
up in the unused port of C$_2$ or in the time bins $t_0'$ or $t_4'$).
The generalization to the case where Alice uses $2N$ time bins to
encode her qubit is immediate. One finds that the visibility becomes
$V_{2N} = {N š/ ( N š- 1 + e^{\sigma^2})}$, see \cite{GLMP}.
\par

The measured average visibilities are plotted in Fig.~\ref{fig2}
as a function of the phase noise standard deviation
in the case where Alice applies $\phi_A=0$ and Bob measures 
the output port P$_1$.
Without noise, the maximal visibilities exceeded 99\%. With unfiltered noise
(lower curve), the visibility decreases exponentially with the amount of noise in accordance with Eq. (\ref{V2}).
The filtration achieved by our setup (upper curve) 
is also very close to the theoretical prediction of Eq.~(\ref{V4}).
Visibilities in the range from 65\% to 78\% are enhanced, by filtration,
to the range from 78\% to 85\%, see inset of Fig.~\ref{fig2}. Noting that
the security threshold $\mathrm{BER}< 11.0$\% translates into $V>78.0$\%
while the insecurity threshold $\mathrm{BER} \ge 14.6$\% corresponds to
$V \le 70.7$\%, we conclude that our setup transforms
a BB84 protocol which is insecure (or of unknown security)
into a provably secure one.  Note that the visibilities were also
tested for the other possible values of $\phi_A = \pi / 2 , \pi ,
3 \pi /2 $ used in the BB84 protocol and for both ports P$_1$ and
P$_2$. The visibilities all exceeded 97.2\% in the case where no
noise is added. The lowest visibilities are
associated with the cases where Alice and Bob choose the
$\{{\pi\over2},{3\pi\over2}\}$ basis since then both parties
must apply a potential to their polarization-sensitive phase
modulator which makes the setup noisier.
\par

In Figure~\ref{fig2}, the visibilities
were measured in a regime where the state, when it leaves Alice's site,
contains approximately 120 photons. In this way, the dark
counts of the detector are negligible. To consider a realistic 
QKD implementation, we also ran the experiment in the single-photon regime,
where the quantum state sent by Alice back to Bob contains approximately
0.8 photons per pulse. The difficulty in this case is that dark
counts become very important (they give a raw error rate of about 30\%)
because we use 3-ns pulses \footnote{By using shorter laser pulses, we should
be able to decrease the dark count rate by at least a factor of 10.}.
Nevertheless, we were able to show in this regime that a visibility
of 69.4\% $\pm 0.2$\% in the absence of filtration can be turned
into 78.8\% $\pm 0.2$\% by filtration, where the statistical error
was calculated for a 95\% confidence level, see open circles in
Fig.~\ref{fig2}. These visibilities are averaged over all four choices 
of $\phi_A$ and for the two output ports $P_1$ and $P_2$, so they are
the relevant quantities for characterizing a QKD scheme. This
averaging explains why these points lie slightly off the curves in
Fig. \ref{fig2}. Note that in QKD the errors due to the dark
counts can in principle be removed using error correction codes
without altering the security, although in practice this is
probably impossible for the high level of dark counts we have here.
\par

We conclude by comparing this method with other QKD schemes that
use higher dimensional systems \cite{CBKG}. The noise model in both
cases is the same: a $d$-dimensional state
$|\psi\rangle$ entering the communication channel exits as
$\rho = e^{-\sigma^2} |\psi\rangle \langle \psi | + ( 1 -
e^{-\sigma^2})\openone/d$. This makes the comparison meaningful. The QKD
schemes based on sets of mutually unbiased bases
considered in \cite{CBKG} can only tolerate a noise level up
to $e^{-\sigma^2}<1/2$. This is because when $e^{-\sigma^2}=1/2$,
there is a simple attack in which the eavesdropper (Eve) 
does not modify the state with probability $1/2$, while,
with probability $1/2$, she keeps the state and sends a random one instead. 
In contrast, error filtration can tolerate arbitrarily
high levels of noise if the dimensionality $d$ of the Hilbert
space is sufficiently large since the visibility $V_d$ tends to
$1$ for large $d$ and fixed $e^{-\sigma^2}$. 
The reason why the above attack no longer works in
the case of error filtration is that when Eve replaces
the quantum state by a random one, the filtration preferentially removes 
the corresponding noisy term. In other words, the noise is replaced 
by a higher effective loss. 
Nevertheless, this also means that a QKD scheme using
error filtration can be vulnerable to attacks which exploit loss.
For instance, if Alice sends attenuated coherent states, Eve could
use a photon-number splitting attack, see \cite{GRTZ} and references 
therein.  In fact, our setup would
probably be insecure against such attacks since we used attenuated
coherent states with 0.8 photons on average and were very close
to the limit of provably secure QKD. This is not a fundamental
problem and could be remedied by further attenuating the states
or by using better approximations of single-photon sources.
\par

In summary, the present experiment demonstrates the optical part 
of a quantum key distribution scheme which can operate with 
a phase-noise level that is too high for a standard implementation 
of the BB84 to be secure. This demonstrates the power and
simplicity of error filtration as a practical method for
circumventing phase noise in quantum communication.
\par

We are pleased to thank N. Gisin, S. Popescu, and N. Linden for
helpful discussions. We acknowledge financial support from the
Communaut\'e Fran\c caise de Belgique under ARC 00/05-251 and from the
IUAP programme of Belgian government under Grant V-18.

\end{document}